\documentclass[runningheads]{llncs}
\usepackage[T1]{fontenc}
\usepackage{marvosym}
\usepackage{graphicx}
\usepackage{comment}
\usepackage{algorithm}
\usepackage{algorithmic}

\begin{document}
\title{A Structure-Aware Irregular Blocking Method for Sparse LU Factorization}
\titlerunning{A Structure-Aware Irregular Blocking Method for Sparse LU Factorization}
%
\author{Zhen Hu \and
Dongliang Xiong\textsuperscript{(\Letter)} \and
Kai Huang \and
Changjun Wu\and
Xiaowen Jiang}
\authorrunning{Z. Hu et al.}
%
\institute{College of Integrated Circuits, Zhejiang University, 311200, China \\
\email{\{huzhen,xiongdl,huangk,wuchangjun,jiangxw\}@zju.edu.cn}}
\maketitle              
\begin{abstract}
In sparse LU factorization, nonzero elements after symbolic factorization tend to distribute in diagonal and right-bottom region of sparse matrices. However, regular 2D blocking on this non-uniform distribution structure may lead to workload imbalance across blocks. Besides, existing matrix features fail to guide us effectively in blocking. In this paper, we propose a structure-aware irregular blocking method for numerical factorization. A novel diagonal block-based feature is introduced to effectively characterize the local nonzero distribution of sparse matrices. Based on this, we further propose an irregular blocking method that adjusts block sizes according to the local distribution of nonzeros. The strategy utilizes fine-grained blocks in dense regions and coarse-grained blocks in sparse regions, adequately balancing the nonzeros of blocks both within the same level and across levels in the dependency tree. Experiments demonstrate that, on a single NVIDIA A100 GPU, our proposed irregular blocking method achieves average speedups of 1.50× and 3.32× over PanguLU and the latest SuperLU\_DIST, respectively. In addition, it achieves speedups of 1.40x and 3.84x over PanguLU and SuperLU\_DIST on 4 NVIDIA A100 GPUs.
\end{abstract}

\section{Introduction}
Solving the sparse linear system \(Ax = b\) ~\cite{davisdirect} is a crucial part of many scientific and engineering problems. There are two ways to solve \(Ax = b\): direct methods that obtain an exact solution in a finite number of steps and iterative methods that obtain an exact solution through continuous iteration. Sparse LU factorization is a commonly used direct method for solving linear systems. It factors the sparse matrix \(A\) of the linear system into a lower triangular matrix \(L\) and an upper triangular matrix \(U\), i.e., \(A = L \times U\). The direct LU factorization~\cite{davis2016survey} method can obtain a more accurate solution and is widely used in scientific computing and simulation.

\begin{figure}[htbp]
\centerline{\includegraphics[width=0.8\linewidth]{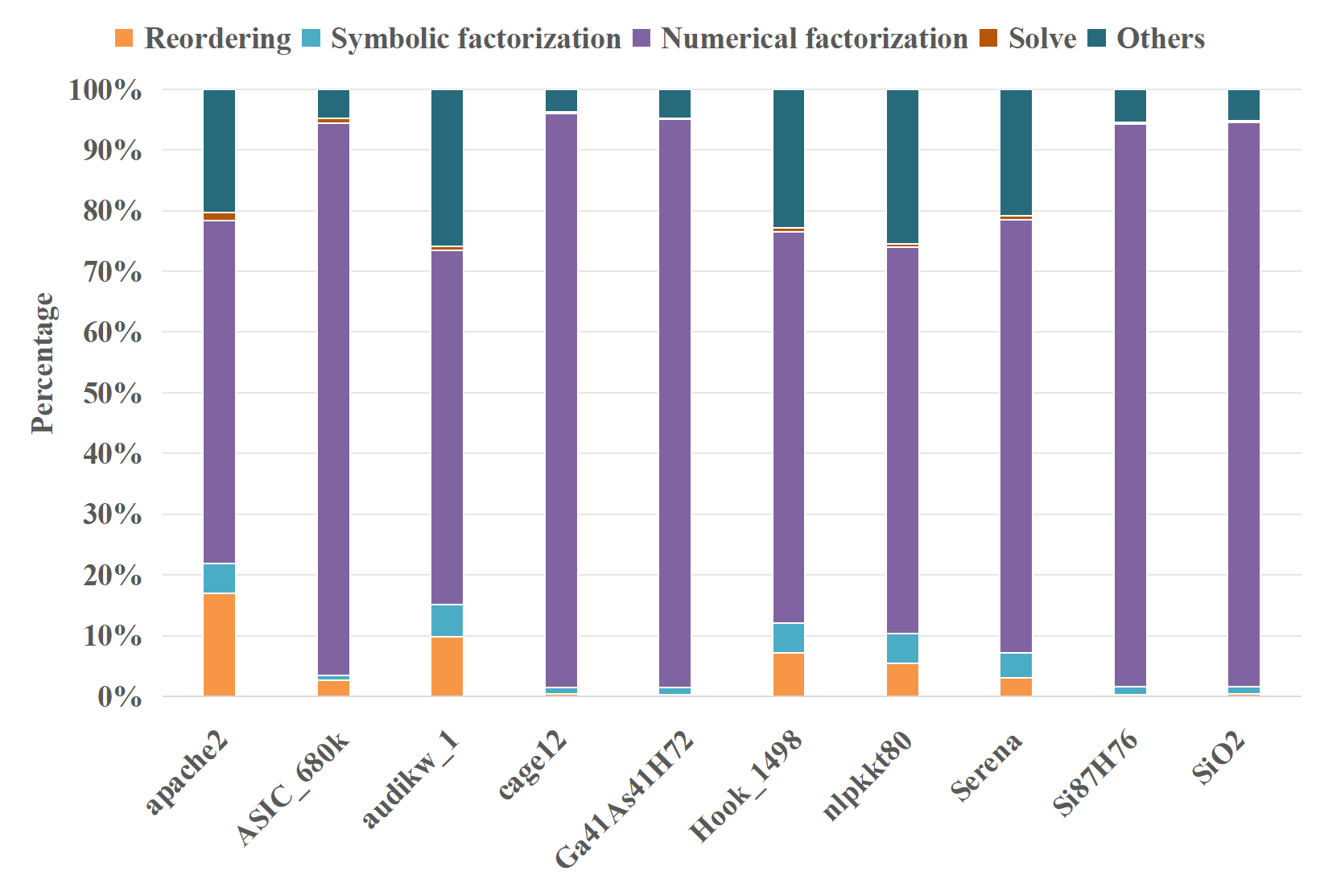}}
\caption{Time breakdown of SuperLU}
\label{breakdown}
\end{figure}
Typical sparse LU factorization is divided into three phases: (1) reordering; (2) symbolic factorization; (3) numerical factorization. The purpose of the reordering phase is to reduce fill-in nonzeros and maintain numerical stability. Then, the symbolic factorization phase is used to determine the nonzero structure of matrices \(L\) and \(U\). Finally, the numerical factorization phase performs floating-point operations to obtain the specific nonzero values. As shown in Fig.~\ref{breakdown}, among these phases, the numerical factorization phase is the most time-consuming, accounting for 50\% to 95\%.
At the same time, there are long dependency chains between tasks in numerical factorization of sparse matrices. A dependency tree~\cite{liu1990role} is a data structure that is used to represent the dependencies between tasks in the factorization of a sparse matrix. 

Preprocessing  sparse matrix to blocks or aggregating similar rows or columns into supernodes~\cite{demmel1999supernodal} can make numerical factorization easier to parallelize. However, existing solvers such as SuperLU\_DIST~\cite{superlu}, UMFPACK~\cite{davis2004umfpack} and PARDISO~\cite{schenk2001pardiso}  exploit dense Basic Linear Algebra Subprograms (BLAS) libraries~\cite{duff2002overview}, which reduce computational efficiency due to excessive fill-ins introduced to aggregate more similar columns.
KLU~\cite{davis2010algorithm} is a direct solver in circuit simulation, based on the Gilbert-Peierls left-looking algorithm. It achieves good performance by permuting the matrices  into Block Triangular Form (BTF), but can only execute factorization in serial. Basker~\cite{booth2016basker} partitions the matrix to a block triangular form where each diagonal block is further partitioned using nested dissection~\cite{khaira1992nested}, generating a number of independent sub-matrices. It performs well in highly sparse matrices, but performs poorly in high density matrices~\cite{chen2024cktso}.

Unlike these solvers, PanguLU~\cite{pangulu} uses a simpler regular two-dimensional blocking method and stores it with a compressed format, then selects either sparse kernels or dense BLAS libraries for computation based on the density of blocks. However, sparse matrices after symbolic factorization demonstrate a structure similar to the Bordered Block Diagonal (BBD) structure~\cite{chen2017parallel}. It features with nonzero elements concentrated in the right-bottom region. Regular 2D blocking will lead to more nonzero elements in blocks near the bottom and right boundaries, and fewer nonzero elements in blocks near the left and upper boundaries. This will cause the last level of the dependency tree during numerical factorization a large computational load and a longer execution time, affecting the overall performance of numerical factorization. Additionally, it will lead to significant differences in the number of nonzero elements between blocks, causing workload imbalance during parallel computation and further impacting overall performance. Therefore, under the premise of sparse storage and computation, efficient preprocessing blocking is crucial for improving numerical factorization performance. Besides, sparse matrices have various structure in real world. Existing matrix features, like density and average nonzeros per row or column, can only provide coarse and one-dimensional information of sparse matrices. PanguLU constructs a selection tree considering matrix dimension and total number of nonzeros after symbolic factorization to decide the block sizes~\cite{jin2024machine}, which fails to select the optimal ones for most sparse matrices.

To mitigate the load imbalance caused by the various non-uniform distribution structure of nonzeros in sparse matrices, we need to address the following two challenges: (1) how to efficiently obtain the nonzero distribution structure of the matrix after symbolic factorization.  (2) how to determine the blocking strategy to alleviate load imbalance.

In this paper, we propose a structure-aware irregular blocking method. Unlike PanguLU blocking sparse matrices into blocks with the same size, our proposed method partitions them adaptively according to the nonzero distribution structure of sparse matrices. The imbalance of nonzeros across blocks can be well alleviated through multiple block sizes.
Firstly, since the sparse matrix after symbolic factorization has a symmetric structure~\cite{bollhofer2020state}, we introduce a diagonal block-based pointer to exploit the two-dimensional distribution characteristics of the sparse matrix. Then, based on the two-dimensional distribution characteristics of the sparse matrix, our proposed blocking method performs coarse-grained blocking in dense regions and fine-grained blocking in sparse regions
to obtain the specific blocking positions. Finally, we conduct experiments on various kinds of test matrices from SuiteSparse Matrix Collection~\cite{davis2010algorithm} on the platform with 4 NVIDIA A100 GPUs. 

The comparisons between our work with some existing relevant solvers is shown in Table~\ref{comparsion} and the contributions of our work are as follows.
\begin{itemize}
    \item A diagonal block-based feature, which is described by percentage of nonzeros along the diagonal, is proposed to characterize global and partial nonzero distribution of sparse matrices. It can be obtained from common compressed formats.
    \item A structure-aware irregular blocking strategy is proposed based on the nonzero distribution of sparse matrices to alleviate workload imbalance of nonzeros.
    \item Experimental results show that our proposed irregular blocking method can achieve significant speedups over PanguLU and the latest SuperLU\_DIST in numerical factorization.
\end{itemize}

\begin{table}[]
\caption{Comparisons between our work with some existing relevant solvers.} \label{comparsion}
\centering
\resizebox{0.8\linewidth}{!}{%
\begin{tabular}{|c|c|c|c|}
\hline
Solver        &  Left or Right looking & Blocking method          & Support GPU or not \\ \hline
KLU           & Left                                                         & Block diagonal                                                      & N                                                            \\ \hline
UMFPACK       & Left                                                        & Multifrontal~\cite{duff1983multifrontal}                                                           & N                                                            \\ \hline
PARDISO    & Left\&Right & Supernode & N \\ \hline
SuperLU\_DIST & Right                                                        & Supernode                                                           & Y                                                           \\ \hline
Basker        & Left                                                         & Recursive Block diagonal & N                                                            \\ \hline
PanguLU       & Right                                                        & Regular 2D block                                                    & Y                                                           \\ \hline

\textbf{Our work}      & \textbf{Right}                                                        & \textbf{Irregular 2D block}                                                  & \textbf{Y}                                                          \\ \hline
\end{tabular}%
}
\end{table}

\section{Background}
\label{sec:background}
\subsection{Sparse LU Factorization}

\begin{figure}[hbp]
\centerline{\includegraphics[width=0.6\linewidth]{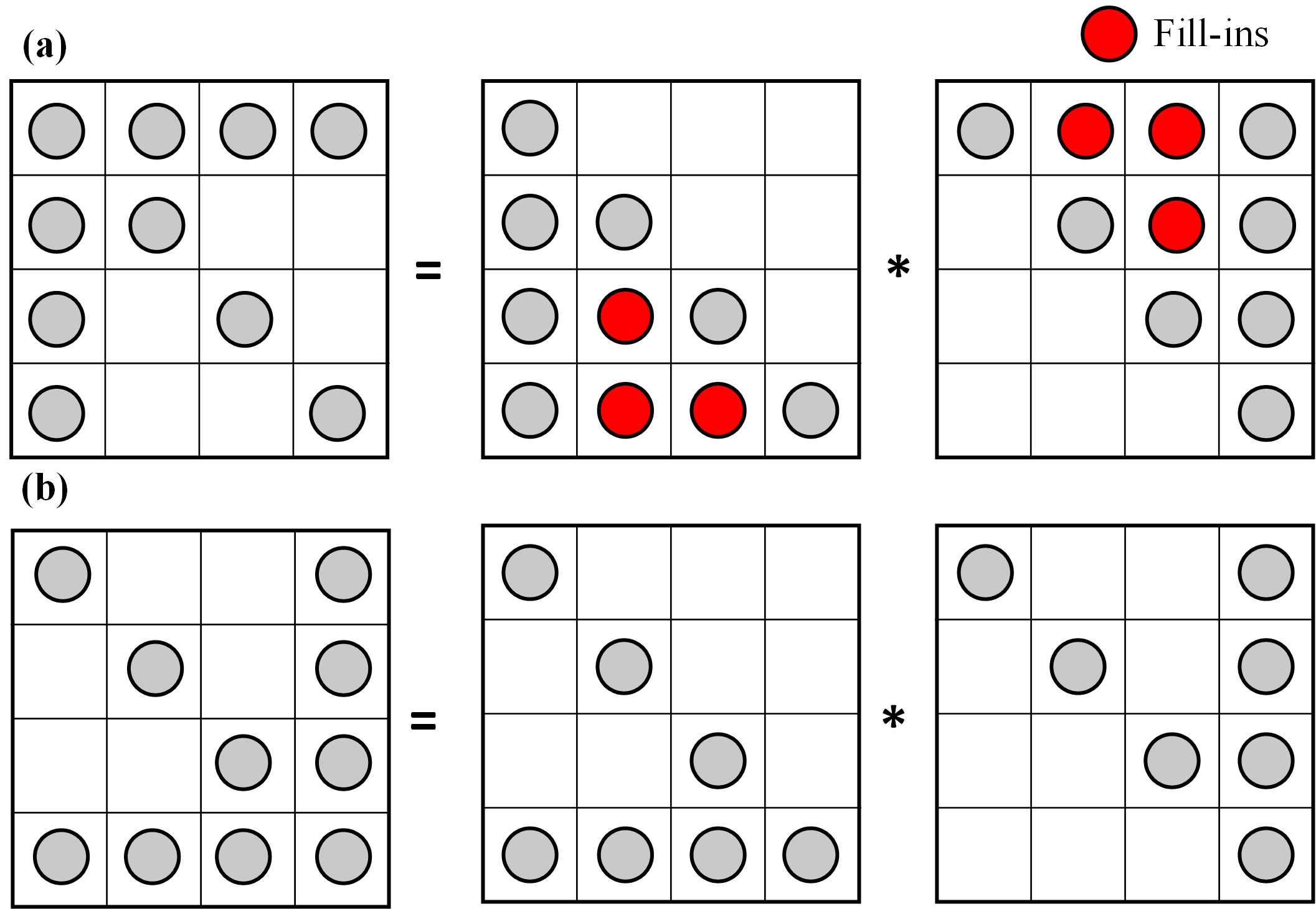}}
\caption{Different structure lead to different fill-ins. (a) A structure leads to full fill-ins. (b) A structure leads to no fill-in.}
\label{fill-ins}
\end{figure}

Sparse LU factorization needs symbolic factorization to predict the symbolic pattern of the LU factors without considering the numerical values. But it will induce extra fill-ins. 
As shown in Fig.~\ref{fill-ins}, in order to minimize fill-ins, reordering algorithms~\cite{demmel1999asynchronous} tend to reorder matrix into a good one just like the below structure~\cite{chen2017parallel}. As a result, after symbolic factorization, most nonzero elements are located in the right-bottom region. 
Besides, blocking technique is an optimization method by exploiting the data locality. Blocked sparse LU factorization is considered except for the extremely sparse matrices.

\subsection{Right-looking Blocked LU Factorization}
\begin{figure}[htbp]
\centerline{\includegraphics[width=\textwidth]{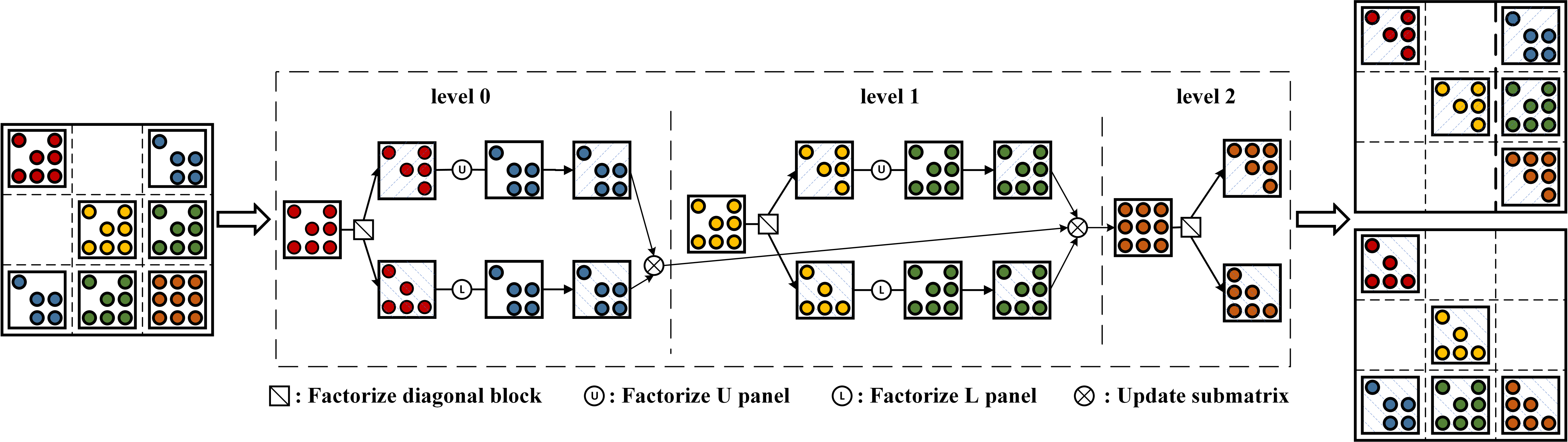}}
\caption{Example of a sparse blocked LU factorization.}
\label{sparse_block_factorization}
\end{figure}

In numerical factorization phase, right-looking LU algorithm is widely used in solvers like SuperLU and PanguLU. The right-looking blocked LU algorithm~\cite{mary2017block}, which treats two-dimensional matrix blocks after preprocessing as input, is described in Algorithm 1. 
These loop nests show that data dependences are frequent. At each level $i$, only after the factorization of diagonal block $B_{i,i}$, row(U) panels $B_{i,*}$ and column(L) panels $B_{*,j}$ have been performed, the entire tailing sub-matrices can be updated. And the $i^{th}$ iteration can only be executed after the $(i-1)^{th}$ iteration have been finished. However, sparsity brings empty blocks which can reduce the computation and data dependency, then introduces more parallelism. 
Computation only need to be performed on non-empty blocks as shown in Fig.~\ref{sparse_block_factorization}. This example adopts regular 2D blocking before numerical factorization and partitions the original matrix into several $3 \times 3$ blocks. And it consists of a dependency tree with $3$ levels.

\begin{algorithm}
\caption{Right-looking Blocked LU Factorization}
\label{alg:example}
\begin{algorithmic}[1]
\REQUIRE a $p \times p$ block matrix B 
\ENSURE $B = [B_{i,j}]_{i=1:p,j=1:p}$
\STATE Construct blocks $B_{i,j}, \forall{i,j}$
\FOR{$i = 1$ to $p$}
    \STATE $B_{i,i}  \to L_{i,i}  U_{i,i}$
    \COMMENT{Factorize the diagonal block}
    \FOR{$j = i+1$ to $p$}
        \STATE $B_{i,j} \gets L_{i,i}^{-1}  B_{i,j}$
        \COMMENT{Factorize U panel}
        \STATE $B_{j,i} \gets B_{j,i} U_{i,i}^{-1}$
        \COMMENT{Factorize L panel}
    \ENDFOR
    \FOR{$j = i+1$ to $p$}
        \FOR{$k = i+1$ to $p$}
            \STATE $B_{k,j} \gets  B_{k,j} - B_{k,i}  B_{i,j}$ 
            \COMMENT{Update submatrix}
        \ENDFOR
    \ENDFOR
\ENDFOR
\end{algorithmic}
\end{algorithm}

\section{Motivation}

\subsection{Ineffective One-dimensional Matrix Features}

\begin{figure}[htbp]
\centerline{\includegraphics[width=0.7\linewidth]{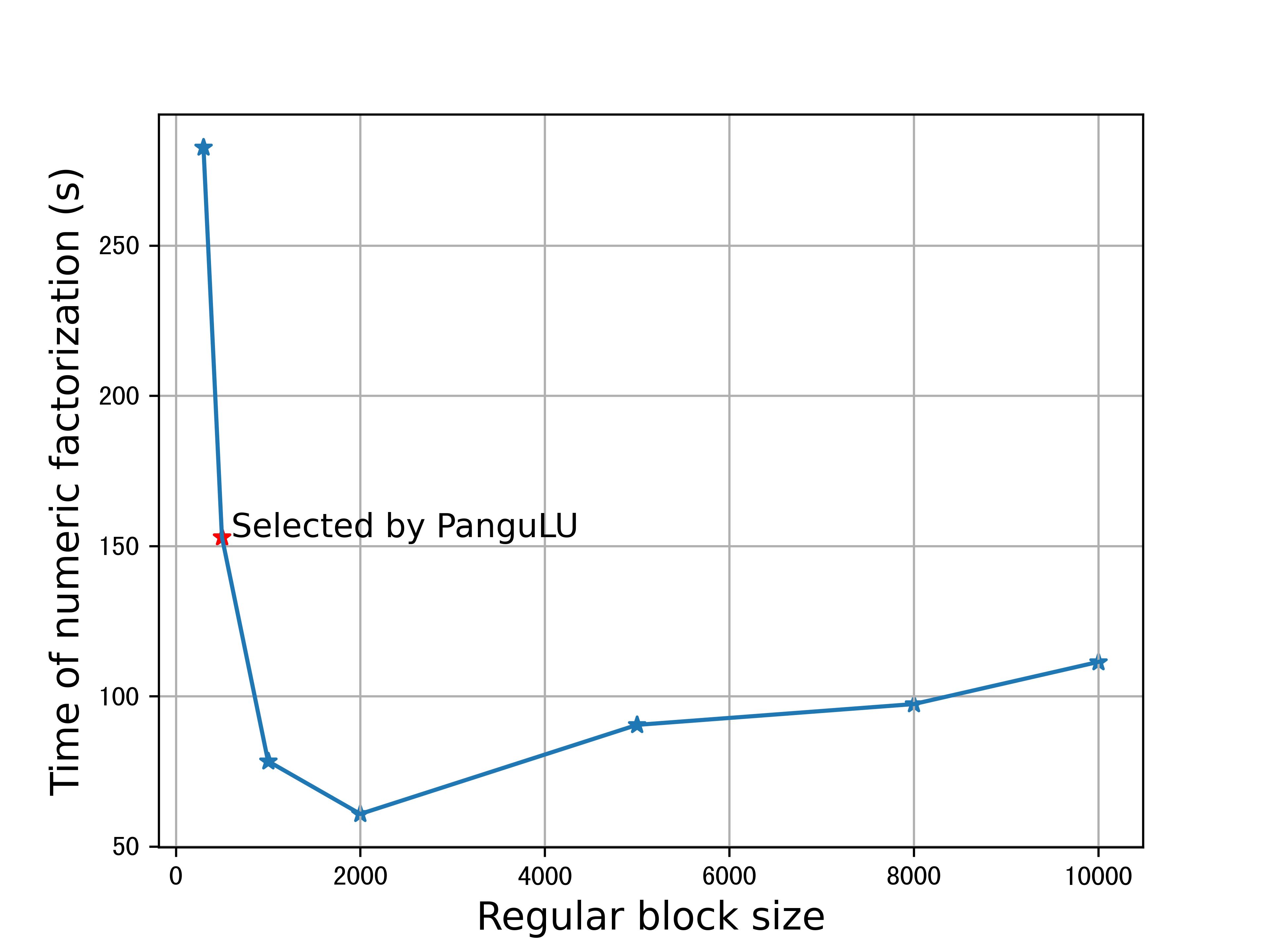}}
\caption{The numerc factorization time varies with different regular block size of a sparse matrix. PanguLU makes choices of block size from 300, 500, 1000, 2000, 5000. But it selects a worse size according to the matrix dimension and density.}
\label{nb}
\end{figure}

The nonzero distribution of a sparse matrix is critical for researchers to accelerate the computation. However, the most common features, like matrix dimension and density, only reflects the features of the whole matrix. Other features, like average number of nonzero elements per row or the standard deviation of nonzero elements in each row, only reflect the one-dimensional feature of the sparse matrix. PanguLU selects a fixed size of regular blocking according to the matrix order and the density of the matrix after symbolic factorization, which can not obtain an optimal block size for the performance of numerical factorization as shown in Fig.~\ref{nb}.
While blocked LU factorization algorithm performs computation with block granularity, it is more important to focus on its two-dimensional features, which have not been explored.

\subsection{Unbalanced Nonzeros}
\begin{figure}[htbp]
\centerline{\includegraphics[scale=0.8]{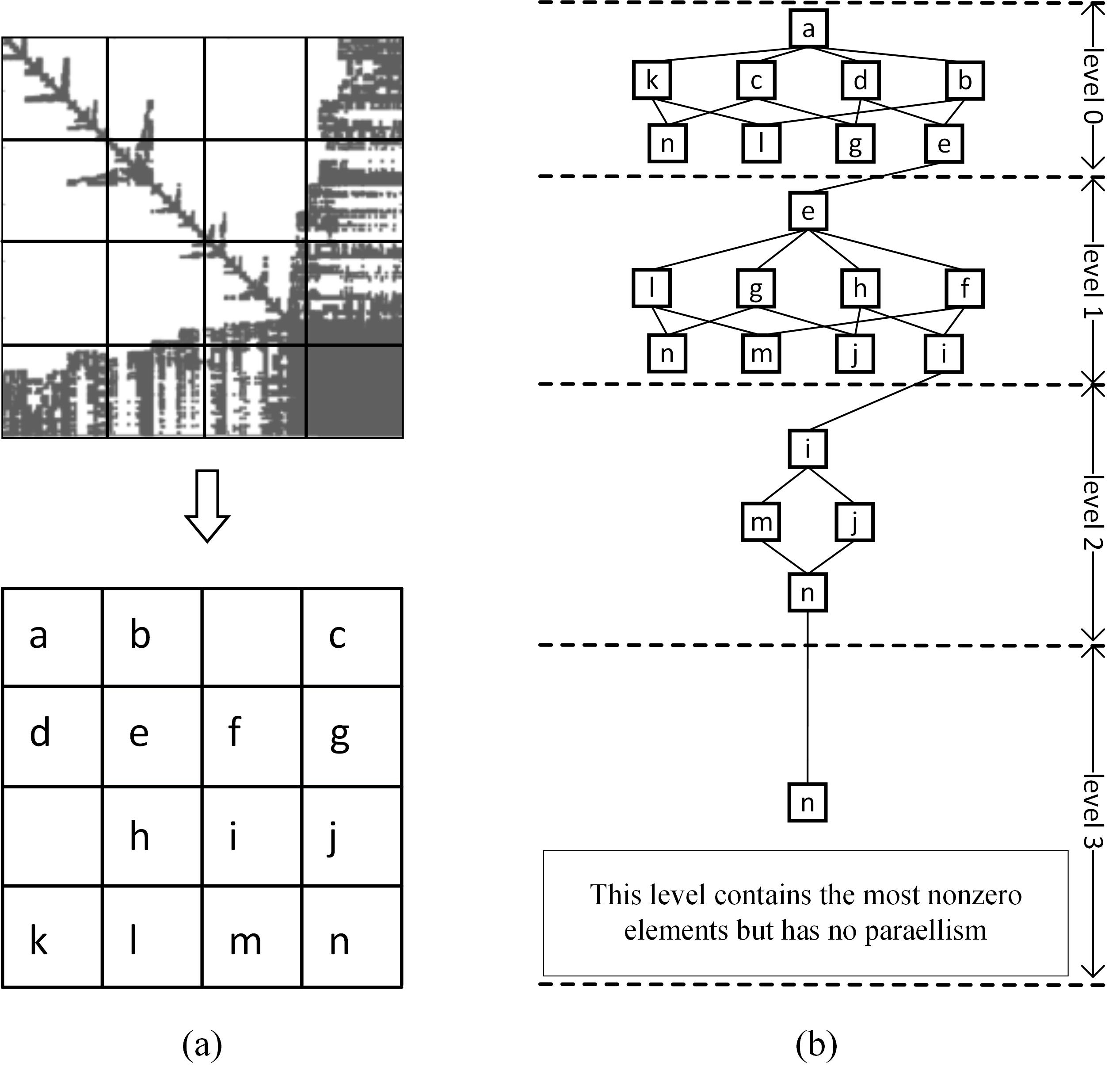}}
\caption{Example of a dependency-level tree of a regularly blocked sparse matrix. (a) Sparse matrix $A$ with regular blocking. (b) The dependency tree of $A$ based on level.}
\label{dependency}
\end{figure}

Regular 2D blocking method partitions matrix into blocks with the same size. However, nonzero elements are distributed non-uniformly, which leads to blocks with unbalanced nonzeros. Besides, since most of nonzero elements locate at the right-bottom region, the density of blocks nearing right and bottom region are generally higher than those near left and up region~\cite{chen2024cktso}. As shown in Fig.~\ref{dependency}, the number of nonzero elements in block $n$ is much more than that in block $a$. This imbalance causes most of the time for numerical factorization to be spent on the last dense blocks. This issue will be more severe in parallel computing as the time difference between blocks induced by load imbalance. Therefore, it is necessary to partition blocks according to the sparsity of region.

\section{Structure-aware Irregular Blocking Method}
\subsection{Overview}
In this paper, we propose a structure-aware irregular blocking preprocessing method for sparse LU numerical factorization. The floating point operations in numerical factorization is related with the number of nonzeros after symbolic factorization. The sparse matrix is stored by Compressed Sparse Column (CSC) format. In the preprocessing phase, we propose a conversion algorithm to get a diagonal block-based pointer from CSC pointer. The diagonal block-based pointer stores the number of nonzeros in diagonal sub-matrix. The percentage of nonzero values distributed along the diagonal can be obtained by dividing the value in the pointer by the total number of nonzero values. To balance the number of nonzeros in blocks, we sample uniformly 1000 points of the nonzeros distribution to guide us in blocking the sparse matrix. The strategy of finer-grained blocks in dense regions while allocating coarser blocks in sparse regions induces blocks with multiple sizes. Workload imbalance can be alleviated through balance the number of nonzeros both within the same level and across levels in the dependency tree.

\subsection{Diagonal Block-based Feature}\label{4.2}
Compressed sparse column (CSC) is a common compressed format for sparse matrix. It utilizes three arrays: column pointer, row index, and value. The column pointer array stores the index in the value array of the first nonzero element of each column. Row index array stores the row index of each nonzero element. And the value array stores the value of nonzero elements of in column order. According to column pointer array, we can get the number of nonzero elements in each column. It can be useful for matrix computation based on columns, like column-wise sparse matrix multiplication~\cite{li2023spada}. Blocked sparse matrix factorization regards blocks as the computational granularity, while compressed sparse column (CSC) only reflect the row distribution of the sparse matrix and cannot effectively guide the performance optimization for blocked matrix factorization.

\begin{figure}[htbp]
\centerline{\includegraphics[width=0.8\linewidth]{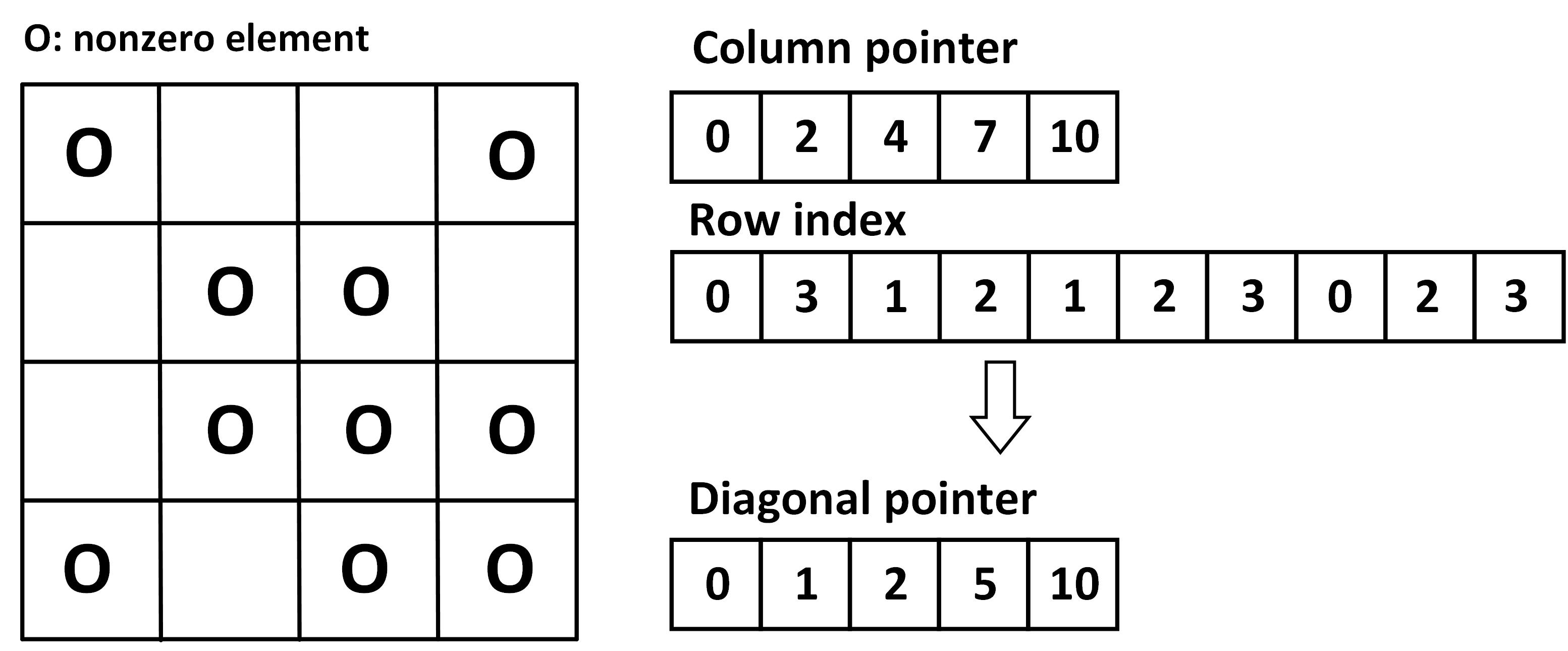}}
\caption{Example of compressed sparse column pointer and diagonal block pointer.}
\label{format}
\end{figure}

Considering the symmetry of the sparse matrix after symbolic factorization~\cite{pangulu}, a diagonal block-based array can be obtained from compressed sparse column pointer, which contains the number of nonzero elements in the sub-matrix \([0:n, 0:n]\), where \(n\) is the row/column index, as shown in Fig.~\ref{format}. The conversion algorithm from arrays of CSC to diagonal block-based pointer is shown in Algorithm~\ref{fc}. 

\begin{algorithm}
\caption{Extracting Diagonal Block Pointer}
\label{fc}
\begin{algorithmic}[1] 
\REQUIRE matrix dimension $n, *columnpter, *rowidx$
\STATE Initialize \(num[n] = 0\)
\FOR{\( i = 0 \) \textbf{to} \( n-1 \)}
    \FOR{\(j = columnptr[i] \) \textbf{to} \(columnptr[i+1] - 1\)}
    \STATE \( index \gets rowidx[j] \)
        \IF{\(index > i\)}
            \STATE{\( num[index] \gets num[index] + 1\)}
        \ENDIF
    \ENDFOR
\ENDFOR
\STATE  \( blockptr[0] \gets 0 \)
\FOR{\( i = 0 \) \textbf{to} \( n-1 \)}
    \STATE{\(num[i] \gets 2 \times num[i] + 1\)}
    \STATE{\(blockptr[i+1] \gets blockptr[i] + num[i]\)}
\ENDFOR
\RETURN \(*blockptr\)
\end{algorithmic}
\end{algorithm}

After normalizing the index and value of diagonal block pointer, we get the feature, i.e., a percentage array of nonzero distribution along the diagonal. When taking them as the \(x\)-\(axis\) and \(y\)-\(axis\), respectively, the resulting curve explicitly shows both global and local distribution features of the sparse matrix.

\begin{figure}[htbp]
\centerline{\includegraphics [width=0.8\linewidth] {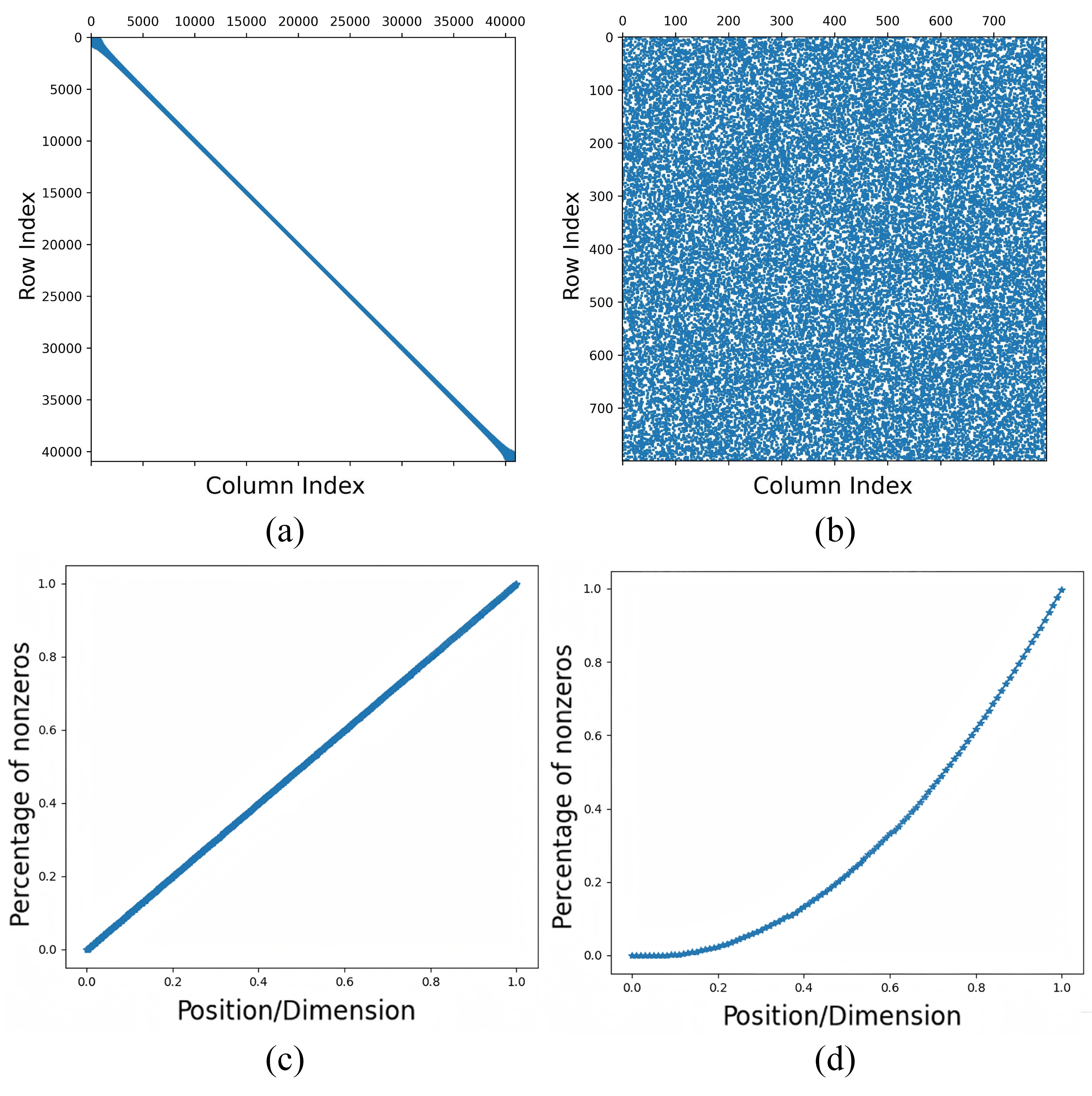}}
\caption{(a) A linear structure matrix. (b) An uniform distribution structure matrix. (c) Percentage of nonzeros based on diagonal pointer of (a). (d) Percentage of nonzeros based on diagonal pointer of (b).}
\label{global}
\end{figure}

\begin{figure}[hbp]
\centerline{\includegraphics [width=0.7\linewidth] {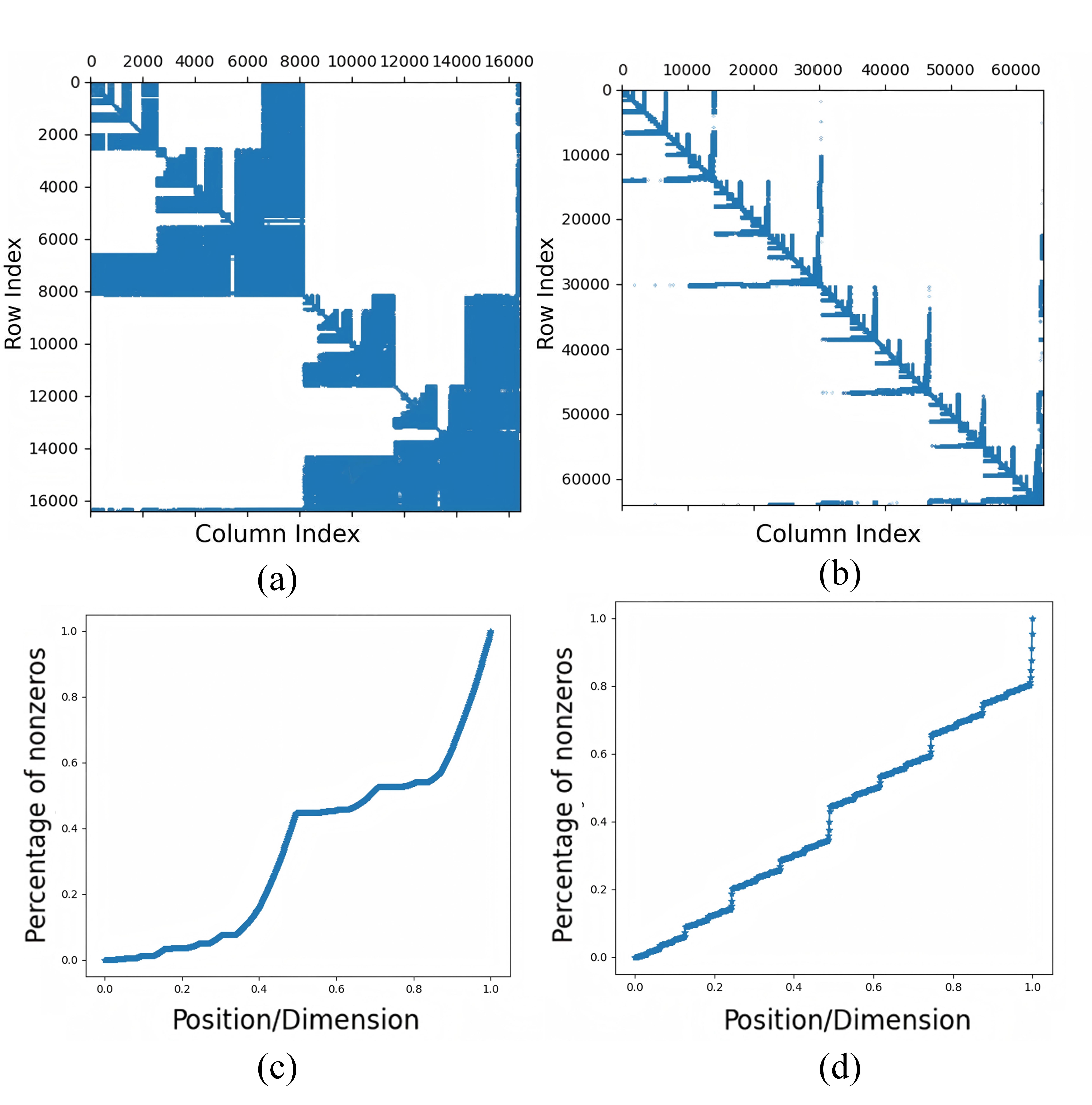}}
\caption{(a) A sparse matrix with some local dense regions. (b) A sparse matrix with some dense rows or columns. (c) Percentage of nonzeros based on the diagonal pointer of (a). 
The partial quadratic trends in the curve indicates that there are local dense regions. 
(d) Percentage of nonzeros based on the diagonal pointer of (b).
The jumps in the curve indicates that there are relatively dense rows and columns.}
\label{partial}
\end{figure}

As shown in Fig.~\ref{global}, for linear matrices, the number of nonzero increases uniformly along the diagonal. Therefore the nonzero percentage curve based on diagonal block pointer shows a linear feature. For uniformly distributed matrices, the number of nonzeros in a block is proportional to the area of the diagonal block. As we can see, the percentage curve of uniformly distributed matrices shows a quadratic feature. We can roughly infer the structure of sparse matrix by the global feature of the curve.
In addition, the local distribution of nonzeros can also be captured. As shown in Fig.~\ref{partial}, figure (a) shows a sparse matrix with some local dense blocks. The nonzero percentage curve based on diagonal block pointer shows partial quadratic trends in dense region with obvious discontinuities. Also, when there are relatively dense rows and columns along the diagonal, there are significant jumps in the curve. The curve in figure (d) shows many jumps because the sparse matrix has relatively dense rows/columns here. Therefore, the diagonal block-based pointer contains the two-dimensional distribution information of a sparse matrix, which can be exploited to guide us in accelerating the block-based matrix computation, like blocked sparse matrix factorization.

\subsection{Irregular Blocking Method Based on Nonzero Distribution}
Considering the locality of the nonzero distribution after symbolic factorization, an irregular blocking method is proposed as described in Algorithm~\ref{blocking algotithm}. The algorithm adopts the strategy of blocking finer-grained blocks for dense regions while allocating coarser blocks for sparse regions.

We sample uniformly 1000 points from the percentage of nonzeros array as input. The algorithm calculate the percentage difference between two consecutive points or points separated by a step interval. If the percentage difference exceeds a threshold, this indicates that the diagonal region between these two indexes contains a significant number of nonzero elements, and the larger index should be marked as a blocking position (\(P_1\)). Otherwise, if the difference does not exceed the threshold, we regard that the diagonal region between the two indexes contains fewer nonzero elements and can be skipped without blocking, while increasing the skip counter $l$. It is hard to set the threshold to distinguish the dense or sparse regions. 
As we have observed in the nonzero percentage curve in Section~\ref{4.2}, the percentage of nonzeros based on diagonal pointer of a matrix with linear structure show a linear feature. Considering the nonzero after symbolic factorization mainly distribute along the diagonal, we regard the linear difference as the threshold, i.e. step/sample\_pointers.
Also, to prevent excessive accumulation of nonzero elements caused by consecutive skips, when $l$ reaches $max\_num$, blocking must still be performed and a blocking position (\(P_m\)) marked. The process described above continues until the index reaches the dimension $N$ of the matrix. The algorithm eventually outputs a blocking position pointer array that records all blocking boundaries. We choose $step$ and $max\_num$ to be 2 and 3 respectively. The parameters are determined empirically based on a large number of parameter tuning experiments.

\begin{algorithm}
\caption{Structure-aware irregular blocking method }
\label{blocking algotithm}
\begin{algorithmic}[1] 
\REQUIRE Percentage array \(pct[sample\_points] \), 
matrix dimension $N$
\STATE \(ptr[0] = 0 \)
\FOR{ index \(i\) = 0  \textbf{to} \( sample\_points \)}
    \IF{\(pct[i+step]-pct[i] \ge threshold\)}
        \STATE \( k \gets k+1 \)
        \STATE \(ptr[k] \gets (i+step)*N/sample\_points \)   \COMMENT{Dense region}
        \STATE \(l \gets 0\)
    \ELSIF{\(l \ge max\_num\)}
        \STATE \( k \gets k+1 \)
        \STATE \(ptr[k] \gets (i+step)*N/sample\_points \)  \COMMENT{Avoid too large blocks}
        \STATE \(l \gets 0\)    
    \ELSE \STATE {\(l \gets l + 1 \)}
    \ENDIF
\ENDFOR
\RETURN blocking position \(*ptr\)
\end{algorithmic}
\end{algorithm}

\begin{figure}[htbp]
\centerline{\includegraphics[width=0.8\linewidth]{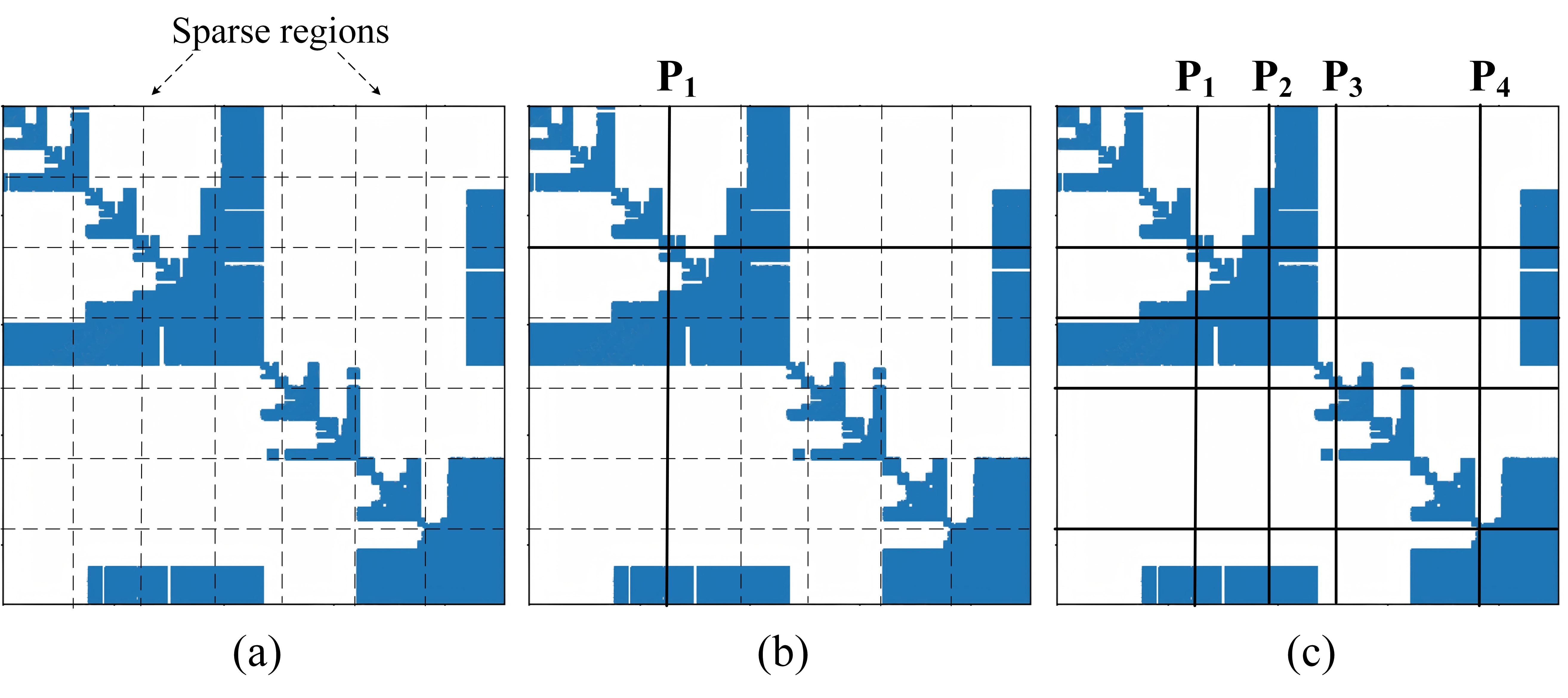}}
\caption{Example of the irregular blocking method. (a) Initial matrix sampled in position marked by dashed lines. (b) Coarse-grained blocking in a sparse region. (c) Final blocking.}
\label{partition example}
\end{figure}

An example is shown in Fig.~\ref{partition example}. We first sample uniformly a proper number of points from the percentage array of nonzero distribution, which means a preliminary regular blocking into some basic blocks. Then according to the difference of distribution percentage between two sample points, we merge a certain number of basic blocks in sparse regions. The final blocking scheme is shown in Fig. 9(c). The blocking positions are denoted by \(P_0\), \(P_1\), ... , \(P_5\), where \(P_0=0\) and \(P_5 = N\).

\section{Experimental Results}
\subsection{Experimental Setup}
We apply the proposed irregular blocking method to PanguLU and conduct experiments on the platform with 4 * NVIDIA A100 GPUs. Hardware information of the test platform is provided in Table~\ref{GPU platform}. SuperLU\_DIST is a parallel extension to the serial SuperLU library. It is targeted for the distributed memory parallel machines. We compile SuperLU\_DIST and PanguLU with gcc-11.3.0, OpenMPI-4.1.4 and CUDA 12.2. Detailed information of the sparse matrices is listed in Table~\ref{benchmark}.  These sparse matrix benchmarks are all downloaded from the SuiteSparse Matrix Collection~\cite{davis2011university}.

\begin{table}[htbp]
\caption{Information of the test platform and baseline algorithm.}\label{GPU platform}
\centering
\resizebox{\linewidth}{!}{%
\begin{tabular}{|c|c|}
\hline
\textbf{GPU Platform} &
  \textbf{Baseline} \\ \hline
\begin{tabular}[c]{@{}c@{}}4 * NVIDIA A100 GPU with 80G, B/W 1555 GB/s \\ Intel Xeon Gold 6348 CPU @2.6GHz \end{tabular} &
  \begin{tabular}[c]{@{}c@{}}PanguLU\\ SuperLU\_DIST 9.1.0\end{tabular} 
   \\ \hline
\end{tabular}%
}
\end{table}

\begin{table}[htbp]
\caption{The  sparse matrices tested in this paper. The matrix order, the number of nonzeros before and after symbolic factorization, the total number of floating-point operations and kinds are listed.}
\centering
\resizebox{0.8\linewidth}{!}{
\begin{tabular}{|c|c|c|c|c|c|}
\hline
\textbf{Matrix}  & \textbf{\begin{tabular}[c]{@{}c@{}}n(A) \\ ($10^{5}$)\end{tabular}} & \textbf{\begin{tabular}[c]{@{}c@{}}nnz(A)\\ ($10^{6}$)\end{tabular}} & \textbf{\begin{tabular}[c]{@{}c@{}}nnz(L+U)\\ ($10^{8}$)\end{tabular}} & \textbf{\begin{tabular}[c]{@{}c@{}}FLOPs\\ ($10^{11}$)\end{tabular}} & \textbf{Kind}                                  \\ \hline
apache2          & 7.15        & 4.82 & 3.19 & 3.46  & Structural Problem      \\ \hline
ASIC\_680k       & 6.83                                                          & 2.64                                                          & 1.18                                                              & 7.68                                                            & Circuit Simulation Problem                     \\ \hline
cage12           & 1.30                                                          & 2.03                                                          & 5.70                                                              & 42.30                                                           & Directed Weighted Graph \\ \hline
CoupCons3D       & 4.17                                                          & 17.28                                                         & 4.99                                                              & 9.05                                                            & Structural Problem                             \\ \hline
dielFilterV3real & 11.03                                                         & 89.31                                                         & 10.77                                                             & 20.08                                                           & Electromagnetics Problem                       \\ \hline
ecology1         & 10.00                                                         & 5.00                                                          & 0.72                                                              & 0.30                                                            & 2D/3D Problem                                  \\ \hline
G3\_circuit      & 15.85                                                         & 7.66                                                          & 1.81                                                              & 0.91                                                             & Circuit Simulation Problem                     \\ \hline
offshore        & 2.60 &       4.24                  & 1.50 &                     1.65                                       & Electromagnetics Problem                             \\ \hline
language & 3.99 & 1.22 & 3.88  & 37.78 & Directed Weighted Graph \\ \hline
boneS10 & 9.15 & 40.88 &  5.29 & 4.86 & Model Reduction Problem \\ \hline
\end{tabular}%
}
\label{benchmark}
\end{table}

\subsection{Performance on A Single GPU}
We first evaluate the performance of the numerical factorization of our work compared with PanguLU and SuperLU\_DIST in a NVIDIA A100 GPU. PanguLU selects the size of block from 200, 300, 500, 1000, 2000 and 5000 according to the order of matrix and number of nonzeros after symbolic factorization. We also run PanguLU with all those block sizes mentioned above, and select the best performance for comparison.
In numerical factorization, both PanguLU and our work use sparse kernels for computation, while SuperLU uses dense kernels. The experimental results in Table~\ref{one_gpu_compared} show that the geometric mean speedups of our work in numerical factorization can achieve 1.50x and 3.32x, compared with PanguLU and SuperLU\_DIST, respectively. 

The sizes of most blocks partitioned by our work are generally larger than those of PanguLU. Large blocks contain relatively more nonzero elements, which can improve the utilization rate of the GPU. 
The speedup over SuperLU\_DIST is mainly due to the computation of sparse kernels.

\begin{table}[htbp]
\caption{The numerical factorization time comparison of SuperLU\_DIST, PanguLU and our work on a NVIDIA A100 GPU}\label{one_gpu_compared}
\resizebox{\textwidth}{!}{
\begin{tabular}{c|c|c|c|c|c}
\hline
\textbf{Matrix}  & \textbf{SuperLU\_DIST(s)} & \textbf{PanguLU(s)} & \textbf{Our work(s)} & \textbf{\begin{tabular}[c]{@{}c@{}}Speedup\\ Our work vs SuperLU\end{tabular}} & \textbf{\begin{tabular}[c]{@{}c@{}}Speedup\\ Our work vs PanguLU\end{tabular}} \\

\hline
apache2          & 26.84          & 11.05      & 8.50        & 3.16x                                                                        & 1.30x                                                                  \\ 
ASIC\_680k       & 84.26          & 70.86      & 16.45       & 5.12x                                                                        & 4.31x                                                                  \\ 
cage12           & 143.44         & 136.41     & 86.45       & 1.66x                                                                        & 1.58x                                                                  \\ 
CoupCons3D       & 26.71          & 32.26      & 17.53       & 1.52x                                                                        & 1.84x                                                                  \\ 
dielFilterV3real & 64.49          & 58.61      & 37.56       & 1.72x                                                                        & 1.56x                                                                  \\ 
ecology1         & 11.49          & 2.41       & 2.37        & 4.84x                                                                        & 1.02x                                                                  \\ 
G3\_circuit      & 26.72          & 5.74       & 4.81        & 5.56x                                                                        & 1.19x                                                                  \\ 
inline\_1        & 29.22          & 7.95       & 7.97        & 3.67x                                                                        & 1.00x                                                                  \\ 
language         & 372.68         & 90.88      & 53.19       & 7.01x                                                                        & 1.71x                                                                  \\ 
boneS10          & 42.94          & 14.30      & 12.90       & 3.33x                                                                        & 1.11x                                                                  \\ \hline
\textbf{GEOMEAN}          & \multicolumn{3}{c|}{}                      & \textbf{3.32x}                                                                        & \textbf{1.50x}
\\ \hline
\end{tabular}%
}
\end{table}

\begin{figure}[htbp]
\centerline{\includegraphics[width=0.8\linewidth]{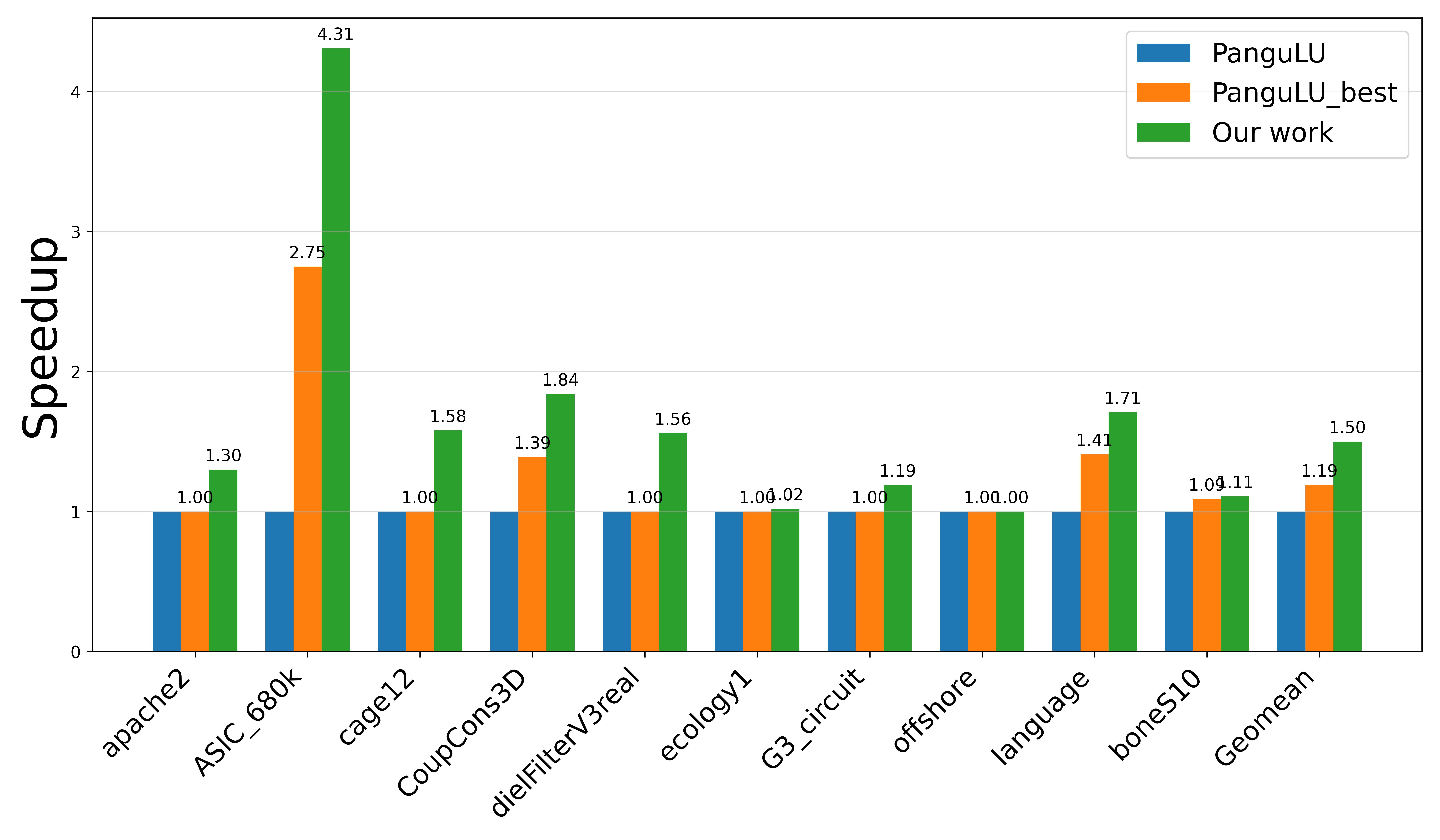}}
\caption{Relative performance speedups on a single A100 GPU.}
\label{gpu1}
\end{figure}

We also test PanguLU with all block sizes in its options and select the best performance for comparison. Fig.~\ref{gpu1} shows the experimental results. PanguLU\_Best means the best performance among options in the selection tree. It achieves an average speedup of 1.19x on the A100 GPU platform over PanguLU, with speedups ranging from 1.0x to 2.75x.
Most of the matrices like apache2 and offshore, can achieve the optimal selection of block size among its options through the selection tree. However, for many matrices such as ASIC\_680k and CoupCons3D, even the best performance achieved through regular blocking remains inferior to that achieved through irregular blocking. The results demonstrate the superiority of our irregular blocking over the regular blocking. 

\subsection{Speedup in  Parallel Computing}
To demonstrate scalability, we also evaluated the performance of numerical factorization with 4 NVIDIA A100 GPUs. The results in Table~\ref{four_gpu_compared} show that our work has a geometric mean of 1.40x compared to PanguLU in the numerical factorization. In addition, the performance achieves a geometric mean of 3.84x compared to SuperLU\_DIST.
In parallel computing, workload imbalance is an important factor that affects performance, especially when dealing with the irregular matrices such as ASIC\_680k. The benefit of workload balance in our work is very obvious, resulting in a more significant speedup. 

\begin{table}[htbp]
\centering
\caption{The numerical factorization time comparison of SuperLU\_DIST, PanguLU and our work on 4 NVIDIA A100 GPUs}\label{four_gpu_compared}
\resizebox{\textwidth}{!}{
\begin{tabular}{c|c|c|c|c|c}
\hline
\textbf{Matrix}  & \textbf{SuperLU\_DIST(s)} & \textbf{PanguLU(s)} & \textbf{Our work(s)} & \textbf{\begin{tabular}[c]{@{}c@{}}Speedup\\ Our work vs SuperLU\end{tabular}} & \textbf{\begin{tabular}[c]{@{}c@{}}Speedup\\ Our work vs PanguLU\end{tabular}} \\
\hline
apache2          & 14.00            & 4.30                           & 3.67                            & 3.82x                                                                  & 1.17x                                                                                      \\
ASIC\_680k       & 26.01            & 18.99                          & 4.66                            & 5.58x                                                                  & 4.08x                                                                                      \\
cage12           & 77.48            & 35.19                          & 23.08                           & 3.36x                                                                  & 1.52x                                                                                      \\
CoupCons3D       & 8.06             & 9.32                           & 5.81                            & 1.39x                                                                  & 1.61x                                                                                      \\
dielFilterV3real & 30.80            & 17.90                          & 16.23                           & 1.90x                                                                  & 1.10x                                                                                      \\
ecology1         & 6.14             & 1.22                           & 1.24                            & 4.96x                                                                  & 0.98x                                                                                      \\
G3\_circuit      & 16.92            & 3.04                           & 2.64                            & 6.42x                                                                  & 1.15x                                                                                      \\
inline\_1        & 15.15            & 4.78                           & 4.43                            & 3.42x                                                                  & 1.08x                                                                                      \\
language         & 177.27           & 25.56                          & 15.60                           & 11.36x                                                                 & 1.64x                                                                                      \\
boneS10          & 21.35            & 8.05                           & 7.17                            & 2.98x                                                                  & 1.12x                                                                                      \\ \hline
\textbf{GEOMEAN}                           & \multicolumn{3}{l|}{}                                             & \textbf{3.84x}                                                                  & \textbf{1.40x}                                                \\                                   
\hline
\end{tabular}%
}
\end{table}

The nonzero distribution of ASIC\_680k and ecology1 are shown below in Fig.~\ref{asic_ecology1}. For ASIC\_680k, we can observe that 98\% of its nonzero elements are located in the bottom and right regions, which are distributed extremely nonuniformly. Our proposed irregular blocking method partitions ASIC\_680K with a large size (about 4000) in sparse regions with only 2\% nonzero elements and a small size (about 1300 and 680) in dense region while PanguLU partitions it with a size of 300 regularly. And there is little data dependence in ASIC\_680k, so our proposed blocking method can achieve a speedup of 4.08x compared with PanguLU. 
For ecology1, the trend of nonzero distribution is almost linear, which means that the nonzeros are distributed along the diagonal uniformly in most region. So our irregular blocking method has no effect in parallel computing. At the same time, our method partitioning ecology1 with larger sizes brings a little reduction in speedup.
Therefore, for those sparse matrices with irregular nonzero distribution, our proposed method can achieve a significant speedup in parallel computing.

\begin{figure}[htbp]
\centerline{\includegraphics[width=\linewidth]{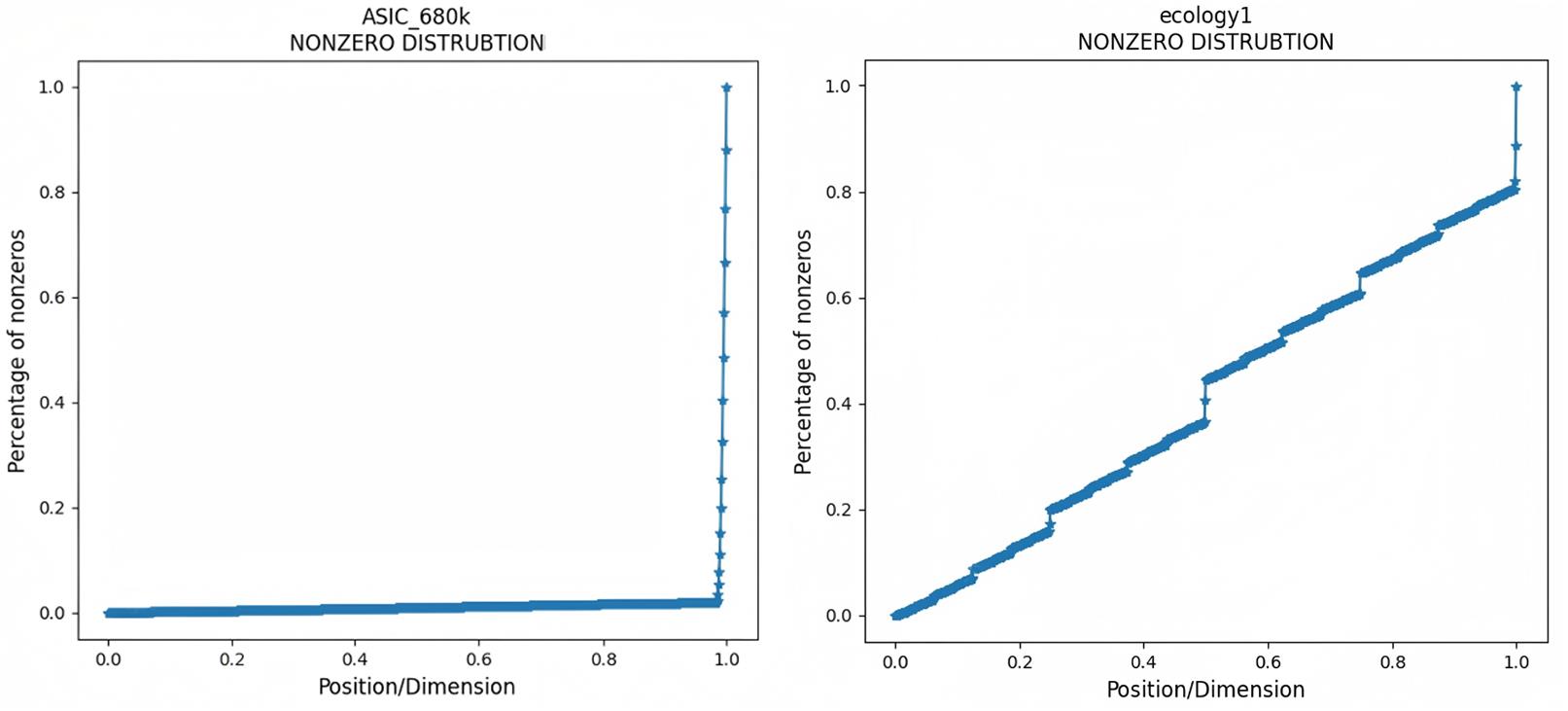}}
\caption{The nonzero distribution of ASIC\_680k (left) and ecology1 (right).}
\label{asic_ecology1}
\end{figure}


\begin{figure}[ht]
\centerline{\includegraphics[width=0.8\linewidth]{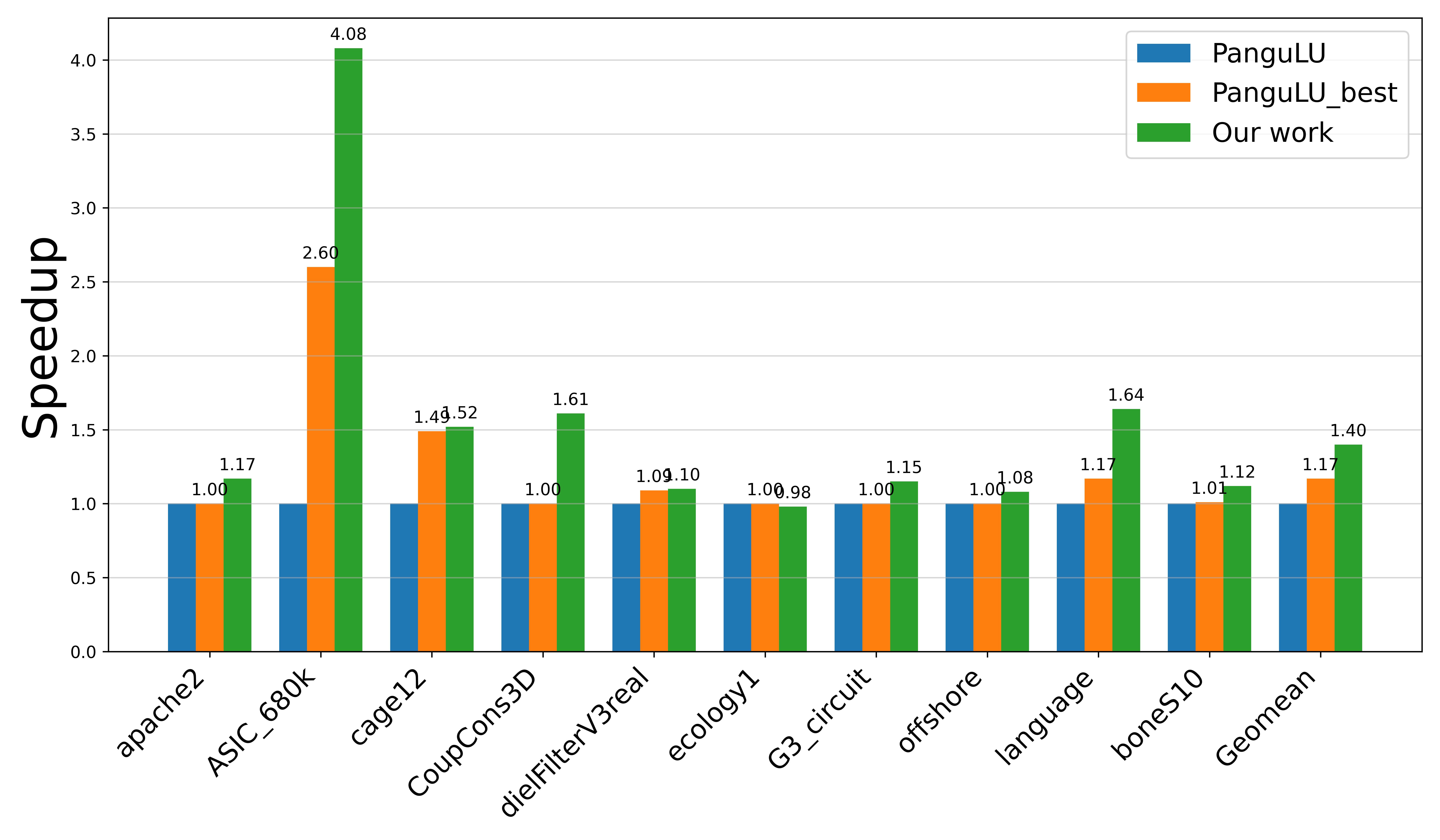}}
\caption{Relative performance speedups on 4 NVIDIA A100 GPUs.}
\label{gpu4}
\end{figure}

We also conduct experiments on PanguLU with all block sizes in its options and select the best performance on 4 NVIDIA A100 GPUs. Fig.~\ref{gpu4} shows the experimental results. PanguLU\_Best achieves an average speedup of 1.17x over PanguLU, with speedups ranging from 1.0x to 2.60x. Our proposed irregular blocking can also achieve better performance than regular blocking. The results in parallel computing show the scalability of our approach and its advantages over regular blocking.

\subsection{Preprocessing Cost}
Regular blocking with the same size in PanguLU makes preprocessing easier and more efficient. Irregular blocking needs to store sizes of block in a size array. The array will be frequently looked up in the preprocessing phase to perform irregular blocking, which causes a longer time of preprocessing. Since the performance in numerical factorization accounts for the largest proportion of the whole LU factorization, it is acceptable to improve the performance of numerical factorization at the cost of increased preprocessing time. 
Besides, optimizing the performance of preprocessing will be our research in the future.

\section{Conclusion}
In this paper, we propose a novel diagonal block-based feature, which is described by percentage of nonzeros along the diagonal, can characterize global and local nonzero distribution of sparse matrices. Furthermore, a structure-aware irregular blocking method is proposed based on the local nonzero distribution. The irregular blocking method adopts the strategy of fine-grained in dense regions and coarse-grained in sparse regions. This strategy can balance the number of nonzero elements between levels in the dependency tree and the number of nonzero elements among blocks within the same level. Experiments on test sparse matrices show that our proposed irregular blocking method achieves speedups of performance in numerical factorization compared with PanguLU and SuperLU\_DIST. On a single NVIDIA A100 GPU, it achieves geometric means of 1.50x and 3.32x, respectively. Also, it achieves speedups of 1.40x and 3.84x over PanguLU and SuperLU\_DIST on 4 NVIDIA A100 GPUs.

%
%

\bibliographystyle{splncs04}
\bibliography{reference}

\end{document}